\documentclass[useAMS,usenatbib]{mnras}

\usepackage{graphicx}
\usepackage{amsmath}
\usepackage{amssymb}
\usepackage{color}
\usepackage{subfig}
\usepackage[usenames]{xcolor}

\newcommand\msun{\, \rm M_\odot}

\newcommand\kms{\, \rm km\,s^{-1}}

\newcommand\aout{{a_{\rm out}}}
\newcommand\ain{{a_{\rm in}}}
\newcommand\vdisp{{v_{\rm disp}}}

\title[S-type planets in close binary stars]{Dynamical origin of S-type planets in close binary stars}
\author[G. Fragione]{Giacomo Fragione$^{1}$\thanks{E-mail: giacomo.fragione@mail.huji.ac.il}\\
$^{1}$Racah Institute for Physics, The Hebrew University, Jerusalem 91904, Israel}

\begin{document}

\maketitle

\begin{abstract}
Understanding the origin of planets that have formed in binary stars is fundamental to constrain theories of binary and planet formation. The planet occurrence rate in binaries with a separation $\lesssim 50$ AU is only $\sim$ one third that of wider binaries or single stars. This may indicate that a close companion has a ruinous influence on planet formation, because it can truncate the protoplanetary disc and pump up planetesimals eccentricity and collision probability. Nevertheless, observations have revealed a few of these systems, which challenge current planet formation theories. Dynamical interactions can deliver planets into S-type orbits. In this paper, we consider as a possible scenario for forming S-type planets in close binaries the single star-binary star interactions that commonly take place in star clusters. We find that the final fraction and orbital properties of S-type planets in close binaries are mainly determined by the mass ratio of the stars involved in the close encounter, and the initial binary and planet semi-major axes. Present and upcoming missions, as TESS, PLATO and CHEOPS may shed new light on the origin of S-type planets in close binaries.
\end{abstract}


\section{Introduction}

The majority of $\gtrsim 1 \msun$ stars are observed in multiple systems, with multiplicity that is function of the stellar mass \citep{duq91,duc13,rag10,tok14a,tok14b}. As a consequence, one of the most generic environments to investigate planet formation and dynamics is that of binary and multiple systems. However, out of $\sim 3700$ confirmed planets\footnote{http://www.exoplanets.org/} only about hundred exoplanets have been found in multiple stellar systems\footnote{http://www.univie.ac.at/adg/schwarz/multiple.html} \citep{rag06,mug09,roe12,sch16}.

The progressive discovery of exoplanets in binary, triple and quadruple stars has opened new frontiers on the study of their origin and stability \citep{naoz13,wang14,thh15,kraus16,frlgin18}. Most of these planets have been observed in binary systems. According to the relative orbital configuration of the binary-planet system, planets in binaries can be divided into S-type, when the planet orbits either of the stars of the binary, and P-type (usually referred to as circumbinary planets), when the planet revolves around both the stars \citep{hagh08}.

The characteristics of planets in binaries have been examined in order to check any possible deviation from the features of planets around single stars. For S-type planets, it is generally believed that the companion star in a wide binary ($a_b\gtrsim 100$ AU) has only a little impact on their formation, and their distribution has been shown to follow the same trend of planets in single stars \citep{desid07,roe12}. The characteristics of exoplanets in close binaries are quite different, and the planet occurrence rate in binaries with a separation $\lesssim 50$ AU is only $\sim$ one third that of wider binaries or single stars. Furthermore, S-type planets have not been found in close binaries with an orbital separation $\lesssim 3$-$5$ AU. This may indicate that the gravitational influence of a close companion can strongly affect \textit{in situ} S-type planet formation and habitability \citep{hagh07,kalt13,wang14,thh15}. Close companions can truncate the S-type protoplanetary disc, thus reducing its mass and lifetime, can pump up planetesimals eccentricity and collision probability, and can make the typical accretion timescale much longer compared to single star systems, so that the protoplanetary disc may dissipate before the planet is formed \citep{theb08,thebm10,xie11,marz12,mirlai15}. Also, the high temperature and vertical turbulence induced by binary perturbations on the disk may prevent planetesimal and then planet formation \citep{nels00,pico13}. 

Although the difficulty in forming planets in close binaries, observations have revealed a few of these systems. In Tab.~\ref{tab:observed}, we report the observed S-type planets in close binaries with separations $a_b\lesssim 50$ AU \footnote{http://exoplanet.eu/planets\_binary/; updated on 11/07/2018} \citep{thh15}. In some cases (marked with "$^{*}$"), we report objects whose mass (or minimum mass) is higher than the conventional $13$ $M_J$ ($M_J$ is Jupiter mass). For most of the systems, only the projected separation between the stars in the binary is known, and the actual orbit remains unconstrained. Also, the relative inclination between the planetary and binary orbital planes is unknown.

Understanding the origin of planets that have formed in these dynamically active environments is crucial to constrain theories of binary and planet formation. Besides \textit{in situ} formation, planets may be delivered into S-type orbits by dynamical processes. \citet{marz07} suggested that some close-binaries hosting planet in S-type orbits may have been part of a former hierarchical triple that became unstable, thus ejecting the third star. In this scenario, the S-type planet formation underwent when the binary pair was farther apart, which was then driven to a tighter orbit during the dynamical instability that ejected the third star. \citet{gong18} discussed that planet-planet scattering among P-type planets, and the subsequent tidal capture, can originate S-type planets in close binaries. Little attention has been devoted to how star interactions can shape the properties of S-type planets. \citet{pfah06} used analytic arguments regarding binary-single and binary-binary scatterings and estimated that dynamical processes can deposit planets in $\lesssim 1\%$ of close binaries. In this paper, we reconsider as a possible scenario for forming S-type planets in close binaries the single star-binary star scatterings that commonly take place in star clusters \citep{lei13}, by means of direct high-precision $N$-body scattering simulations. Our planet is originally bound to a single star that has the chance to interact with a binary star during its journey in the cluster environment. We consider different masses for the stars and the planet, and study the role of the orbital semi-major axis and eccentricity of both the planet and of the binary star. Finally, we also study the effect of the planet mass and velocity dispersion on the dynamical formation of S-type planets in close binaries.

\begin{table}
\caption{Observed close binary stars with S-type planets: name, planet semi-major axis ($a_{p}$), binary semi-major axis ($a_b$).}
\centering
\begin{tabular}{lcc}
\hline
Name & $a_{p}$ (AU) & $a_b$ (AU) \\
\hline
\hline
Kepler 693      & $0.11\ $   & $2.9$  \\
Kepler 420      & $0.38\ $   & $5.3$  \\
HD59686         & $1.09\ $   & $13.6$ \\
OGLE2013BLG0341 & $0.70\ $   & $15.1$ \\
HD7449          & $2.29\ $   & $18.1$ \\
HD87646         & $0.12\ $   & $19.6$ \\
                & $1.57^{*}$ & $19.6$ \\
HD41004A        & $1.63\ $   & $20.2$ \\
HD41004B        & $0.02^{*}$ & $20.2$ \\
Gliese 86       & $0.11\ $   & $21.1$ \\
HD196885        & $2.58\ $   & $21.1$ \\
HD8673          & $2.99^{*}$ & $35.3$ \\
HD164509        & $0.87\ $   & $36.8$ \\                
K2-136          & $0.12\ $   & $40.2$ \\
WASP-11         & $0.04\ $   & $42.4$ \\
OGLE2008BLG092L & $14.9\ $   & $48.2$ \\
\hline
\end{tabular}
\label{tab:observed}
\end{table}

\begin{table*}
\caption{Models: mass of the stars in the binary ($m_1=m_2$), mass of the planet-host star ($m_3$), planet's mass ($m_P$), binary semi-major axis ($a_b$), planet semi-major axis ($a_{pl}$), binary eccentricity ($e_b$), planet eccentricity ($e_{pl}$), velocity dispersion ($\vdisp$).}
\centering
\begin{tabular}{lcccccccccc}
\hline
Name & $m_1=m_2$ ($\msun$) & $m_3$ ($\msun$) & $m_p$ (M$_J$) & $a_b$ (AU) & $a_p$ (AU) & $e_b$ & $e_p$ & $\vdisp$ ($\kms$) \\
\hline
\hline
Model 1   & $1$;$3$;$5$ & $1$ & $1$ & $10$;$20$;$50$;$100$ & $1$ & $0$ & $0$ & $3$ \\
\\
Model 2   & $1$;$3$;$5$ & $1$ & $1$ & $10$ & $1$ & $0$;$0.2$;$0.4$;$0.6$ & $0$ & $3$ \\
\\
Model 3   & $1$;$3$;$5$ & $1$ & $1$ & $10$ & $1$ & $0$ & $0$;$0.2$;$0.4$;$0.6$ & $3$ \\
\\
Model 4   & $1$;$3$;$5$ & $1$ & $1$ & $10$ & $0.1$;$0.5$;$1$ & $0$ & $0$ & $3$ \\
\\
Model 5   & $1$;$3$;$5$ & $1$ & $0.3$;$1$;$10$ & $10$ & $1$ & $0$ & $0$ & $3$ \\
\\
Model 6   & $1$;$3$;$5$ & $1$ & $1$ & $10$ & $1$ & $0$ & $0$ & $0.3$;$1$;$3$;$5$  \\
\\
\hline
\end{tabular}
\label{tab:models}
\end{table*}

The paper is organised as follows. In Section 2, we describe the initial conditions and the numerical method we used in our simulations. In Section 3, we present the results of our $N$-body scattering experiments, and estimate the number of close-binaries hosting an S-type planet that are produced in open and globular clusters, in Section 4. Finally, in Section 5, we draw our conclusions and discuss the implications of our findings.

\section{Numerical Simulations}

Our numerical experiments were performed principally using \textsc{fewbody}, a numerical code suited for simulating small-$N$ gravitational dynamics \citep{fregeau04}.

We summarise the initial conditions of our runs in Table \ref{tab:models}. The binary star is made up of equal-mass stars ($m_1=m_2$), with masses of $1$-$3$-$5 \msun$. We consider initial semi-major axis $a_b$ in the range $10$ AU-$100$ AU, and initial eccentricities $e_b=0$-$0.6$. The planet-host stars is taken to be $m_3=1\msun$ star, while the planet has mass ranging from Saturn mass ($m_p=0.3$ M$_J$) to Super-Jupiter mass ($m_p=10$ M$_J$). The semi-major axis of the planet ($a_p$) is in the interval $0.1$ AU-$1$ AU, and its initial eccentricity $e_p=0$-$0.6$. Finally, we fix the relative velocity of the binary star and the planet-host star to the velocity dispersion of the host cluster, which we vary in the range $\vdisp=0.3$-$5\kms$, typical of open clusters and globular clusters. The impact parameter is drawn from a distribution
\begin{equation}
f(b)=\frac{b}{2b_{\rm max}^2}\ .
\end{equation}
In the previous equation, $b_{\rm max}$ is the maximum impact parameter of the scattering experiment defined by
\begin{equation}
b_{\rm max}=p_{\rm max}\sqrt{1+\frac{2G M_T}{p_{\rm max} v_{\rm disp}^2}}\ ,
\label{eqn:bmax}
\end{equation}
where $M_T=m_1+m_2+m_3+m_P$ is the total mass of the system and $p_{\rm max}$ is the maximum pericentre distance of the encounter. We set $p_{\rm max}=5(a_b+a_p)$ \citep{heg96}.

Four angles describe the relative orientations and phases of the binary star encounter with the planet-host star. Given the plane of motion of the binary and the planet-host star, the relative inclinations of the orbital plane of the binary and of the planet's orbital plane constitute the first set of two angles. Finally, the initial relative phases of the stars in the binary and of the planet along its orbit around the host star add two more angles. For all our scattering experiments, these angles are chosen randomly. 

The initial separation of the binary and planet-host star is chosen to be the distance at which the tidal perturbation on both systems has a fractional amplitude $\delta=F_{\rm tid}/F_{\rm rel}=10^{-5}$. In the previous equation, $F_{\rm rel}$ and $F_{\rm tid}$ are the relative force and the initial tidal force between each component of the two systems, respectively \citep{fregeau04,antogn16}. \textsc{fewbody} classifies the results of the scattering events into a set of independently bound hierarchies and considers a run completed when their relative energy is positive and the tidal perturbation on each outcome system is smaller than $\delta$. When a triple system, as in the case of a planet orbiting a star in a binary, is formed, \textsc{fewbody} checks its stability through the \citet{mar01} stability criterion
\begin{equation}
\frac{\aout}{\ain}>\frac{2.8}{1-e_{\rm out}}\left[\left(1+\frac{m_{\rm out}}{m_{\rm in}}\right)\frac{1+e_{\rm out}}{\sqrt{1-e_{\rm out}}}\right]^{2/5}\left(1-\frac{0.3 i}{180^\circ}\right)\ .
\label{eqn:macrit}
\end{equation}
In the previous equation, $m_{\rm in}=m_{\rm in,1}+m_{\rm in,2}$ is the total mass of the inner system made up of the planet and its host star, while $m_{\rm out}$ is the mass of the companion star and $e_{\rm out}$ the binary star orbital eccentricity. Finally, $i$ is the relative inclination between the planet orbital plane and the binary orbital plane.

\section{Results}

\begin{figure} 
\centering
\includegraphics[scale=0.55]{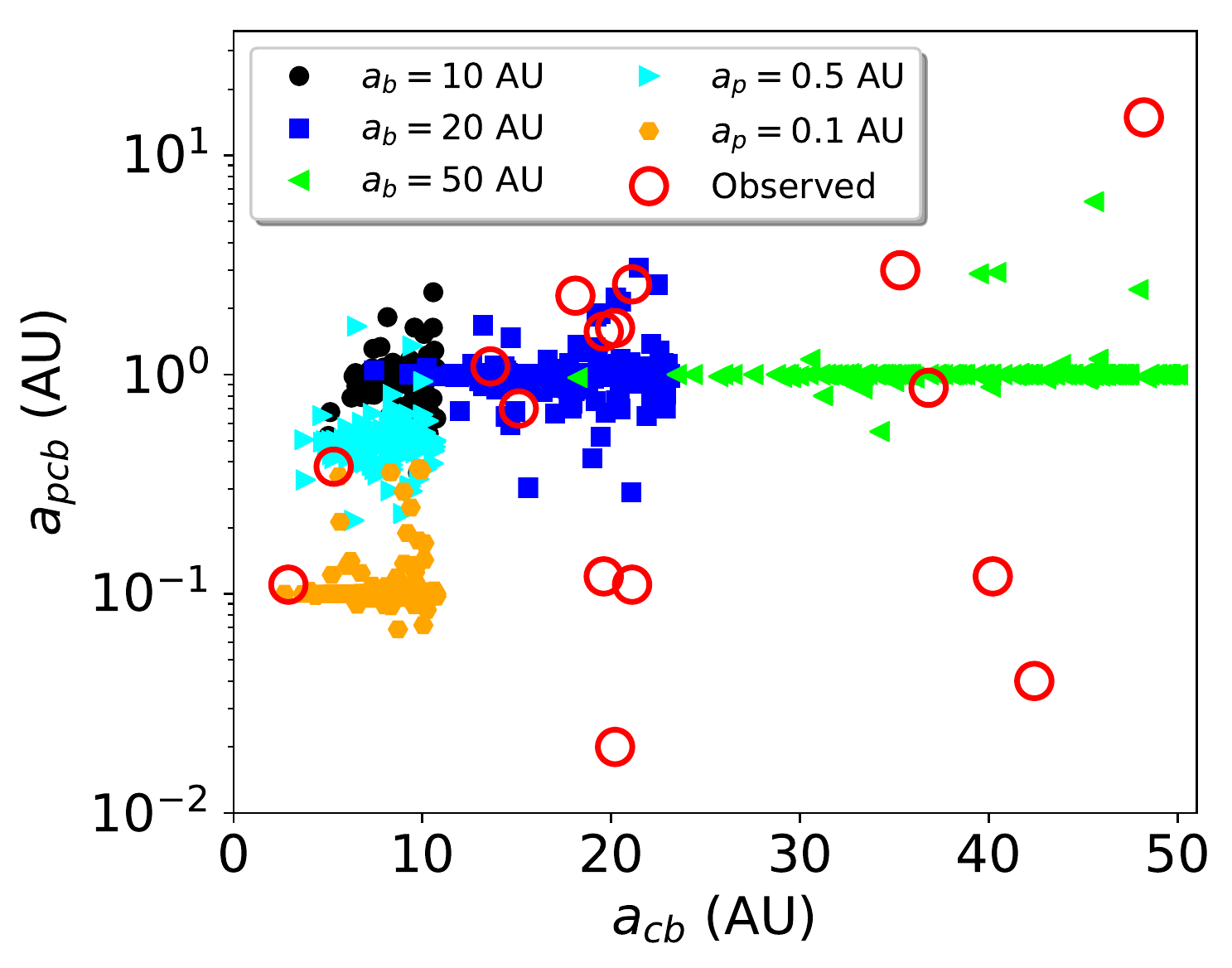}
\caption{Final distribution of the semi-major axis $a_{\rm cb}$ of the dynamically formed close binaries, that host an S-type planet, as function of the planet orbital semi-major axis $a_{\rm pcb}$ for $m_1=m_2=1 \msun$ from Model 1 ($a_b=10$-$20$-$50$-$100$ AU) and Model 5 ($a_p=0.1$-$0.5$-$1$ AU). Shown also the observed S-type planets in close binary stars (from Tab.~\ref{tab:observed}).}
\label{fig:semimod15}
\end{figure}

\begin{table*}
\caption{Results: mass of stars in the binary ($m_1=m_2$), mass of the planet-host star ($m_3$), planet's mass ($m_P$), binary semi-major axis ($a_b$), planet semi-major axis ($a_p$), binary eccentricity ($e_b$), planet eccentricity ($e_p$), velocity dispersion ($\vdisp$), final fraction of binaries that host a planet ($f_{bp}$), final fraction of close binaries that host a planet ($f_{cbp}$).}
\centering
\begin{tabular}{cccccccccccc}
\hline
$m_1=m_2$ ($\msun$) & $m_3$ ($\msun$) & $m_p$ (M$_J$) & $a_b$ (AU) & $a_p$ (AU) & $e_b$ & $e_p$ & $\vdisp$ ($\kms$) & $f_{\rm bp}$ (\%) & $f_{\rm cbp}$ (\%)\\
\hline
\hline
$1$ & $1$ & $1$   & $10$  & $1$ & $0$ & $0$ & $3$ & $2.15$ & $2.15$ \\
$1$ & $1$ & $1$   & $20$  & $1$ & $0$ & $0$ & $3$ & $4.31$ & $4.31$ \\
$1$ & $1$ & $1$   & $50$  & $1$ & $0$ & $0$ & $3$ & $6.12$ & $1.74$ \\
$1$ & $1$ & $1$   & $100$ & $1$ & $0$ & $0$ & $3$ & $6.55$ & $0.00$\\
$3$ & $1$ & $1$   & $10$  & $1$ & $0$ & $0$ & $3$ & $0.22$ & $0.22$ \\
$3$ & $1$ & $1$   & $20$  & $1$ & $0$ & $0$ & $3$ & $0.27$ & $0.27$ \\
$3$ & $1$ & $1$   & $50$  & $1$ & $0$ & $0$ & $3$ & $0.51$ & $0.39$ \\
$3$ & $1$ & $1$   & $100$ & $1$ & $0$ & $0$ & $3$ & $0.41$ & $0.07$\\
$5$ & $1$ & $1$   & $10$  & $1$ & $0$ & $0$ & $3$ & $0.38$ & $0.38$ \\
$5$ & $1$ & $1$   & $20$  & $1$ & $0$ & $0$ & $3$ & $0.79$ & $0.79$ \\
$5$ & $1$ & $1$   & $50$  & $1$ & $0$ & $0$ & $3$ & $0.77$ & $0.71$ \\
$5$ & $1$ & $1$   & $100$ & $1$ & $0$ & $0$ & $3$ & $0.33$ & $0.00$ \\
$1$ & $1$ & $1$   & $10$ & $1$ & $0.2$ & $0$ & $3$ & $2.06$ & $2.06$ \\
$1$ & $1$ & $1$   & $10$ & $1$ & $0.4$ & $0$ & $3$ & $2.04$ & $2.04$ \\
$1$ & $1$ & $1$   & $10$ & $1$ & $0.6$ & $0$ & $3$ & $1.73$ & $1.73$ \\
$3$ & $1$ & $1$   & $10$ & $1$ & $0.2$ & $0$ & $3$ & $0.17$ & $0.17$ \\
$3$ & $1$ & $1$   & $10$ & $1$ & $0.4$ & $0$ & $3$ & $0.23$ & $0.23$ \\
$3$ & $1$ & $1$   & $10$ & $1$ & $0.6$ & $0$ & $3$ & $0.16$ & $0.16$ \\
$5$ & $1$ & $1$   & $10$ & $1$ & $0.2$ & $0$ & $3$ & $0.30$ & $0.30$ \\
$5$ & $1$ & $1$   & $10$ & $1$ & $0.4$ & $0$ & $3$ & $0.31$ & $0.31$ \\
$5$ & $1$ & $1$   & $10$ & $1$ & $0.6$ & $0$ & $3$ & $0.21$ & $0.21$ \\
$1$ & $1$ & $1$   & $10$ & $1$ & $0$ & $0.2$ & $3$ & $2.28$ & $2.28$ \\
$1$ & $1$ & $1$   & $10$ & $1$ & $0$ & $0.4$ & $3$ & $1.83$ & $1.83$ \\
$1$ & $1$ & $1$   & $10$ & $1$ & $0$ & $0.6$ & $3$ & $1.82$ & $1.82$ \\
$3$ & $1$ & $1$   & $10$ & $1$ & $0$ & $0.2$ & $3$ & $0.19$ & $0.19$ \\
$3$ & $1$ & $1$   & $10$ & $1$ & $0$ & $0.4$ & $3$ & $0.16$ & $0.16$ \\
$3$ & $1$ & $1$   & $10$ & $1$ & $0$ & $0.6$ & $3$ & $0.16$ & $0.16$ \\
$5$ & $1$ & $1$   & $10$ & $1$ & $0$ & $0.2$ & $3$ & $0.29$ & $0.29$ \\
$5$ & $1$ & $1$   & $10$ & $1$ & $0$ & $0.4$ & $3$ & $0.20$ & $0.20$ \\
$5$ & $1$ & $1$   & $10$ & $1$ & $0$ & $0.6$ & $3$ & $0.30$ & $0.30$ \\
$1$ & $1$ & $1$   & $10$ & $0.1$ & $0$ & $0$ & $3$ & $8.56$ & $8.56$ \\
$1$ & $1$ & $1$   & $10$ & $0.5$ & $0$ & $0$ & $3$ & $4.53$ & $4.53$ \\
$3$ & $1$ & $1$   & $10$ & $0.1$ & $0$ & $0$ & $3$ & $0.49$ & $0.49$ \\
$3$ & $1$ & $1$   & $10$ & $0.5$ & $0$ & $0$ & $3$ & $0.30$ & $0.30$ \\
$5$ & $1$ & $1$   & $10$ & $0.1$ & $0$ & $0$ & $3$ & $0.63$ & $0.63$ \\
$5$ & $1$ & $1$   & $10$ & $0.5$ & $0$ & $0$ & $3$ & $0.58$ & $0.58$ \\
$1$ & $1$ & $0.3$ & $10$ & $1$ & $0$ & $0$ & $3$ & $2.08$ & $2.08$ \\
$1$ & $1$ & $10$  & $10$ & $1$ & $0$ & $0$ & $3$ & $2.43$ & $2.43$ \\
$3$ & $1$ & $0.3$ & $10$ & $1$ & $0$ & $0$ & $3$ & $0.20$ & $0.20$ \\
$3$ & $1$ & $10$  & $10$ & $1$ & $0$ & $0$ & $3$ & $0.27$ & $0.27$ \\
$5$	& $1$ & $0.3$ & $10$ & $1$ & $0$ & $0$ & $3$ & $0.33$ & $0.33$ \\
$5$ & $1$ & $10$  & $10$ & $1$ & $0$ & $0$ & $3$ & $0.32$ & $0.32$ \\
$1$	& $1$ & $1$   & $10$ & $1$ & $0$ & $0$ & $0.3$ & $1.68$ & $1.68$ \\
$1$	& $1$ & $1$   & $10$ & $1$ & $0$ & $0$ & $1$   & $2.25$ & $2.25$ \\
$1$	& $1$ & $1$   & $10$ & $1$ & $0$ & $0$ & $5$   & $2.34$ & $2.34$ \\
$3$	& $1$ & $1$   & $10$ & $1$ & $0$ & $0$ & $0.3$ & $0.33$ & $0.33$ \\
$3$	& $1$ & $1$   & $10$ & $1$ & $0$ & $0$ & $1$   & $0.21$ & $0.21$ \\
$3$	& $1$ & $1$   & $10$ & $1$ & $0$ & $0$ & $5$   & $0.17$ & $0.17$ \\
$5$	& $1$ & $1$   & $10$ & $1$ & $0$ & $0$ & $0.3$ & $0.36$ & $0.36$ \\
$5$	& $1$ & $1$   & $10$ & $1$ & $0$ & $0$ & $1$   & $0.30$ & $0.30$ \\
$5$	& $1$ & $1$   & $10$ & $1$ & $0$ & $0$ & $5$   & $0.35$ & $0.35$ \\
\\
\hline
\end{tabular}
\label{tab:results}
\end{table*}

We consider in total 6 different models, as summarised in Table \ref{tab:models}. We ran 50k scattering experiments for each set of initial conditions for a total of $\sim 2.7$ million simulations.

In our models, we consider different mass ratios $q=2m_{3}/(m_{1}+m_{2})$ by fixing the mass of the planet-host star to $1\msun$ and varying the masses of the stars in the binary (assumed to be equal-mass) in the range $1-5\msun$. In Model 1, we study the effect of the initial binary semi-major axis, by considering $a_b$ in the range $10$-$100$ AU, while fixing both binary and planet eccentricities $e_b=e_p=0$, and the velocity dispersion $\vdisp=3\kms$. We also set the planet mass and semi-major axis to $m_p=1$ M$_J$ and $a_p=1$ AU, respectively. In Model 2, we analyse the effect of the initial binary eccentricity ($e_b=0-0.6$), while in Model 3, we investigate the influence of the planet initial eccentricity ($e_b=0-0.6$). We study the effect of the initial planet semi-major axis by varying it in the range $a_p=0.1-1$ AU in Model 4, and study the role of the planet mass ($m_p=0.3$-$10$ M$_J$) in Model 5. Finally, in Model 6, we investigate the effect of the velocity dispersion by considering the range $\vdisp=0.3-5\kms$.

Ignoring the planet, there are three possible fates of a binary star-single star encounter: (i) the binary is disrupted as a consequence of the fly-by of the single star; (ii) the binary remains bound on an a perturbed orbit; (iii) the single star replaces one of the stars in the binary, ejecting the former companion. During the encounter, the planet may become unbound with respect to the system, remain bound to the initial host star or exchange host. Table~\ref{tab:results} reports all the different parameters of all the runs considered in this work, along with the relative rates (in percent) of dynamically formed binaries ($f_{\rm bp}$) and close binaries ($f_{\rm cbp}$; binary semi-major axis $\le 50$ AU) with an S-type planet. The final rates depend almost only on $a_b$, $a_p$ and $q$. The initial binary semi-major axis largely determines the success rate of dynamically formed binaries with a planet. Typically, this happens when one of the two stars in the binary is exchanged with the planet-host star. For $q=1$, binaries with initial semi-major axis of $10$ AU, $20$ AU, $50$ AU, and $100$ AU produce dynamically planet-host binaries in $\sim 2.15$\%, $\sim 4.13$\%, $\sim 6.12$\%, and $\sim 6.55$\% of the encounters. However, while in the first two cases all the binaries are close binaries, only $\sim 20\%$ of them have final semi-major axis $\le 50$ AU in the case of $a_b=50$ AU. These results are a factor of a few larger than the previous analytical estimates by \citet{pfah06}. We find almost no dynamically-formed close binaries with a planet when $a_b=100$ AU. Smaller initial planet semi-major axes produce a larger number of systems. Actually, the planet is likely ejected during a close encounter since it is the less massive object in the interaction. The mass ratio also determines the number of dynamically formed binaries, which for $q<1$ is typically a factor of $\sim 5$-$10$ times smaller than the equal-mass case. In these cases, the planet likely becomes unbound during the close interaction with the massive binary. The initial binary and planet eccentricities, the planet mass and the velocity dispersion do not affect significantly the final rates.

\subsection{Planet and binary semi-major axis distributions}

\begin{figure} 
\centering
\includegraphics[scale=0.55]{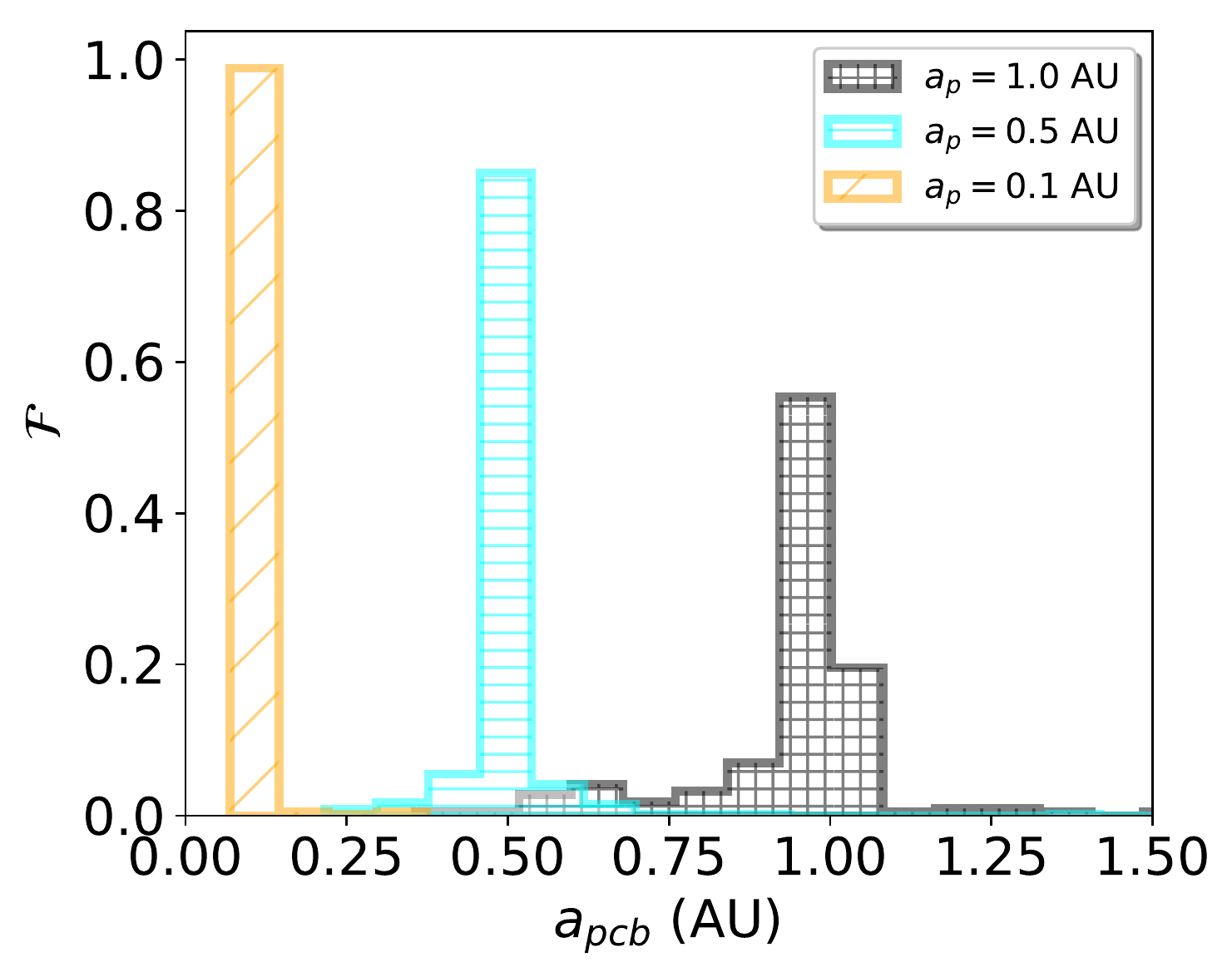}
\caption{Semi-major axis distributions of the S-type planets in close binaries in Model 4 for $m_1=m_2=1 \msun$ (mass ratio $q=1$). Planet orbital semi-major axes are typically distributed around a central value that corresponds to the initial planet semi-major axis $a_p$.}
\label{fig:apcbmod4}
\end{figure}

Figure \ref{fig:semimod15} shows the final distribution of the semi-major axis $a_{\rm cb}$ of the dynamically formed close binaries, that host an S-type planet, as function of the planet orbital semi-major axis $a_{\rm pcb}$. We report the final $a_{\rm cb}$ and $a_{\rm pcb}$ for $m_1=m_2=1 \msun$ (mass ratio $q=1$) from Model 1 ($a_b=10$-$20$-$50$-$100$ AU) and Model 5 ($a_p=0.1$-$0.5$-$1$ AU). We also plot for reference the current observed S-type planets in close binary stars (Tab.~\ref{tab:observed}). Different regions of the $a_{\rm cb}$-$a_{\rm pcb}$ plane are populated by close binaries formed from binaries with different initial semi-major axis. The larger the initial $a_b$, the broader the semi-major axis distribution. We find that the typical semi-major axis of a close binary that hosts an S-type planet is comparable to the initial binary semi-major axis, when $a_b<50$ AU. This corresponds to the cases where all the final planet-host binaries are close-binaries. In the case $a_b=50$ AU, only a few of the planet-host binaries are close binaries. The unequal-mass cases ($q<1$) are qualitatively similar in the plane $a_{\rm cb}$-$a_{\rm pcb}$, but the final number of dynamically formed close binaries with a planet is a factor of $\sim 5$-$10$ times smaller. The location on the $a_{\rm cb}$-$a_{\rm pcb}$ plane is also set by the initial planet semi-major axis. Figure \ref{fig:apcbmod4} illustrates that the semi-major axes of dynamically formed S-type planets, $a_{\rm cb}$, are typically distributed around a central value that corresponds to the initial planet semi-major axis. If a close binary that host an S-type planet was formed dynamically as a consequence of a close encounter between a binary star and a planet-host single star, the plane $a_{\rm cb}$-$a_{\rm pcb}$ offers an important tool to understand and trace its origin.   

\subsection{Planet eccentricity}

\begin{figure} 
\centering
\includegraphics[scale=0.55]{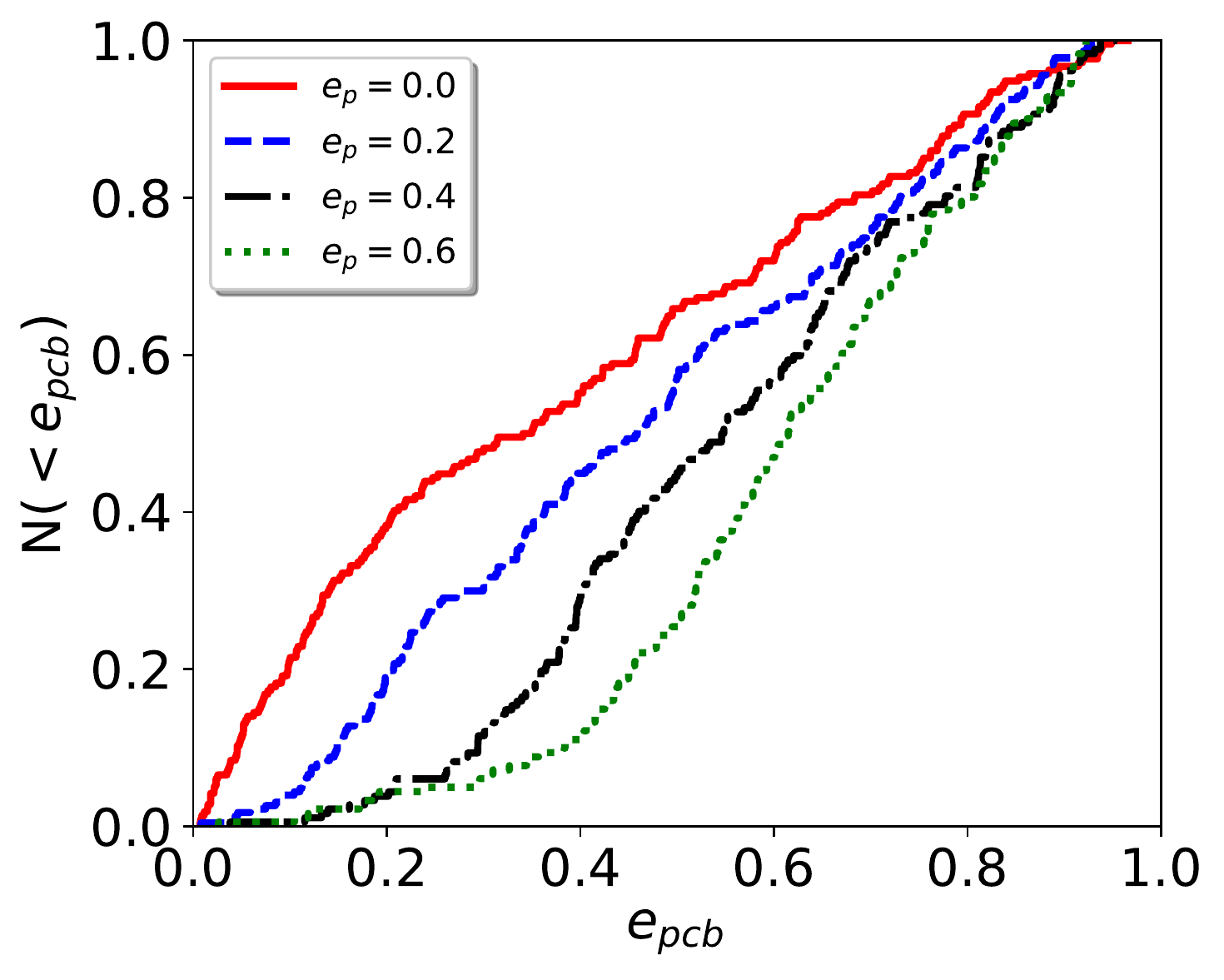}
\caption{Eccentricity of the S-type planets in close binaries in Model 3 for $m_1=m_2=1 \msun$ (mass ratio $q=1$). The initial eccentricity of the planet shapes the eccentricity of the S-type planets: the larger $e_p$, the larger the typical $e_{\rm pcb}$.}
\label{fig:eccplmod3}
\end{figure}

As discussed, only the mass ratio $q$, the initial binary semi-major axis $a_b$, and the initial planet semi-major axis $a_p$ play a role in determining the final fraction of dynamically formed S-type planets in close binaries. While the initial planet eccentricity does not affect the fraction of formed systems, it influences the final eccentricity of the planet. Figure~\ref{fig:eccplmod3} shows the cumulative distribution of the eccentricity of the S-type planets in close binaries in Model 3 for $m_1=m_2=1 \msun$ (mass ratio $q=1$). The initial eccentricity of the planet shapes the eccentricity of the S-type planets: the larger $e_p$, the larger the typical $e_{\rm pcb}$. While $\sim 60$\% of the planets have $e_{\rm pcb}\lesssim 0.4$ when $e_p=0$, this fraction decreases to $\sim 10$\% in the case $e_p=0.6$.

\subsection{Planet inclination}

\begin{figure} 
\centering
\includegraphics[scale=0.55]{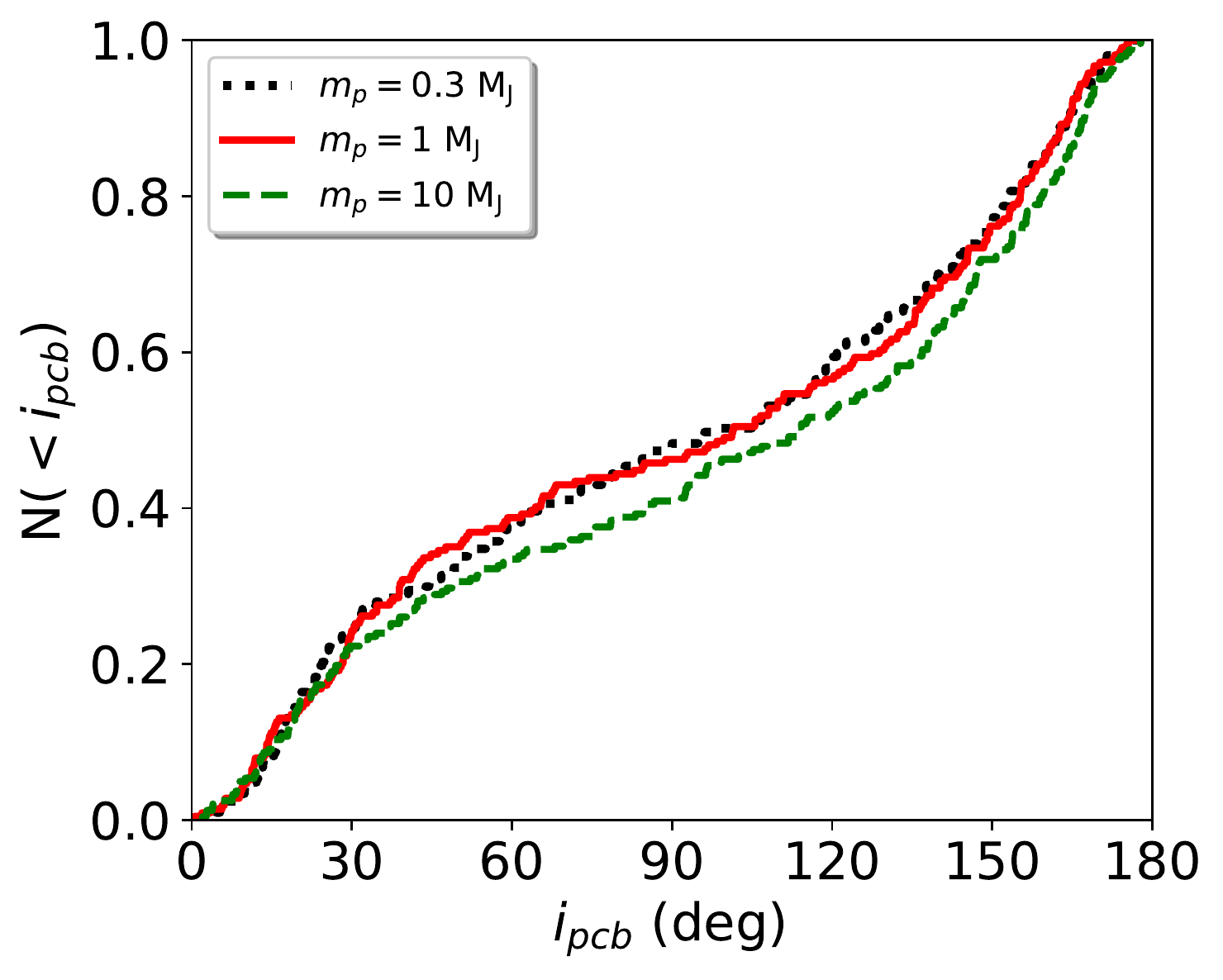}
\caption{Orbital inclination of the S-type planets (with respect to the host binary orbital plane) in close binaries in Model 5 for $m_1=m_2=1 \msun$ (mass ratio $q=1$).}
\label{fig:incmod5}
\end{figure}

The relative inclination of the S-type planets is a fundamental quantity, still unconstrained, that affects the probability of observing these planets \citep{mart17}. Figure~\ref{fig:incmod5} shows the cumulative distribution of the orbital inclination of the S-type planets, with respect to the host binary orbital plane, in close binaries in Model 5 for $m_1=m_2=1 \msun$ (mass ratio $q=1$). The inclination distribution is nearly independent of the planet mass, and is also independent of the other parameters taken into account in this work. We note that there are a few systems within $\sim 40$ deg-$140$ deg. This angle window corresponds to the Kozai-Lidov window, where the conservation of the angular momentum of the 3-body system dictates oscillations in the planet orbital eccentricity and inclination \citep{koz62,lid62}. As a consequence of the variations in the eccentricity, which may reach almost unity for highly inclined systems, the planet can hit the host star and merge. Hence, these systems are almost unstable (see Eq.~\ref{eqn:macrit}), and S-type planets cannot exist in these configurations.

\section{Formation rates}

The parameter space of the scattering events considered in this work is very large, and various parameters intervene in determining the final probability of dynamically forming S-type planets in close binaries. Having a precise estimate of the relative importance of the mechanism we have discussed is practically difficult, and deserves future work. Nevertheless, we can use the results of our numerical scattering simulations as a proxy to roughly estimate how often binary-single encounters leading to the formation of a close-binary hosting an S-type planet may have occurred within star clusters.

We follow \citet{port05} and evaluate the overall encounter rate as 
\begin{equation}
\Gamma\approx f_{s} f_{sp} f_{b} N_* n \sigma \vdisp\ ,
\end{equation}
where $N_*$ and $n$ are the number of star within the cluster and its density, respectively, $f_{s}$ and $f_{b}$ are the single and binary fraction of the cluster, respectively, and $f_{sp}$ is the fraction of single stars that host a planet. The latter quantity depends both on the planet mass and size, and on the physical properties of the host star, as its metallicity \citep{fress13,hsu18,nara18}. For a large variety of systems, $f_{sp}\sim 1$-$50$\% \citep{win15}. The cross-section of our scattering experiments is computed from
\begin{equation}
\sigma=\pi b_{\max}^2 f_{\rm cbp}\ ,
\end{equation}
where $b_{\max}^2$ is calculated from Eq.~\ref{eqn:bmax} and $f_{\rm cbp}$ from the results of our scattering experiments (last column Tab.~\ref{tab:results}). The total number of events within a time $T$ in $N_C$ clusters can therefore be estimated as
\begin{equation}
N\approx \Gamma T =f_{s} f_{sp} f_{b} N_* N_C n \sigma \vdisp T\ .
\end{equation}
After substitution
\begin{eqnarray}
N\approx 630 \left(\frac{f_{s}}{0.5}\right) \left(\frac{f_{sp}}{0.2}\right) \left(\frac{f_{b}}{0.5}\right)\left(\frac{N}{1000}\right)\left(\frac{N_C}{1000}\right)\times \nonumber\\
\times \left(\frac{n}{1\ \mathrm{pc}^{-3}}\right) \left(\frac{\sigma}{5 \times 10^4\ \mathrm{AU}^2}\right) \left(\frac{\vdisp}{1\ \mathrm{km\ s}^{-1}}\right) \left(\frac{T}{10\ \mathrm{Gyr}}\right)\ .
\end{eqnarray}
Here, we have scaled the result to the typical density, number of stars and velocity dispersion of open clusters \citep{spz2010}. The cross-section has been normalized to the average cross-section (in the case $a_b=10$ AU) we have computed from our scattering experiments, for all Solar-like equal-mass stars ($m_1=m_2=m_3=1\msun$). Finally, we have scaled to the typical number of open clusters observed in the Milky Way \citep{spz2010}. Our calculations show that more than $\sim 600$ close-binaries with a planet in S-type orbit may have been formed through the process studied in this paper in the open cluster system of our Galaxy ($\sim 1$-$2$ per cluster), assuming a constant number of open clusters throughout the Milky Way's history. As well, the same process can occur in the core of globular clusters, where the density, number of stars and velocity dispersion are larger than the typical open clusters \citep{harr16}, thus leading to a number of systems $\sim 1000$ times larger.

\section{Discussion and Conclusion}

In this paper, we have considered as a possible scenario for forming S-type planets in close binaries the single star-binary star scatterings that commonly take place in star clusters. Our planet is originally bound to a single star that interacts with a binary star during its journey in the cluster environment. We have studied different stellar and planetary masses, different orbital semi-major axis and eccentricity of the planet and of the binary star, and different velocity dispersion. We have found that the three main parameters that determine the final fraction and orbital properties of S-type planets in close binaries are the mass ratio of the stars involved in the close encounter, the binary semi-major axis, and the initial planetary orbital semi-major axis. The initial planet eccentricity does not affect the final numbers, but only the final eccentricity of the S-type planet. The other parameters (planet mass, velocity dispersion and initial binary eccentricity) do not play a role in this scenario. 

We have then used the results of our numerical simulations as a proxy to estimate how many close-binaries hosting an S-type planet may have been formed in the Milky Way's open clusters and globular clusters, as a consequence of single-binary scatterings. We have evaluated that $\sim 600$ systems may have been formed in open clusters, while the number of systems possibly originated in the cores of globular clusters can be as high as $\sim 10^5$.

Understanding the origin of planets that have formed in binary stars is fundamental to constrain theories of binary and planet formation. If S-type systems in close binaries cannot be formed \textit{in-situ}, they should be mainly the result of planet trapping either as a consequence of the evolution of the stars of a binary or as a consequence of encounters, both of which ought to be rare. Both the rarity of these event and the difficulty in forming these planets \textit{in-situ} seem to agree with the fact that most of the campaigns for detecting these planets have either produced no results, or the number of their discovered planets did not exceed one or two. The recently launched TESS, along with future data by the James Webb Telescope and upcoming exoplanets missions like PLATO and CHEOPS, may shed new light on the origin of S-type planets in close binaries.

\section*{Acknowledgements}
We thank Nader Haghighipour for useful and insightful comments, and the anonymous referee for a constructive discussion. GF is supported by the Foreign Postdoctoral Fellowship Program of the Israel Academy of Sciences and Humanities. GF also acknowledges support from an Arskin postdoctoral fellowship and Lady Davis Fellowship Trust at the Hebrew University of Jerusalem. Simulations were performed on the \textit{Astric} cluster at the Hebrew University of Jerusalem.

\bibliographystyle{mn2e}
\bibliography{biblio}
\end{document}